\documentstyle[iopfts]{ioplppt}

\begin{document}
\title{The geometry of the Barbour-Bertotti theories I. The reduction process}
\author{L\'{a}szl\'{o} \'{A} Gergely \dag\ddag}
\address{\dag Laboratoire de Physique Th\'{e}orique,
Universit\'{e} Louis Pasteur,
3-5 rue de l'Universit\'{e} 67084 Strasbourg, France}
\address{\ddag KFKI Research Institute for Particle and Nuclear Physics,
Budapest 114, P.O.Box 49, H-1525 Hungary}
 
\begin{abstract}
The dynamics of $N\geq 3$ interacting particles is investigated in the
non-relativistic context of the Barbour-Bertotti theories. The reduction
process on this constrained system yields a Lagrangian in the form of a
Riemannian line element. The involved metric, degenerate in the flat
configuration space, is the first fundamental form of the space
of orbits of translations and rotations (the Leibniz group). The Riemann
tensor and the scalar curvature are computed by a generalized Gauss formula
in terms of the vorticity tensors of generators of the rotations.
The curvature scalar is further given in terms of the principal moments of
inertia of the system. Line configurations are singular for $N\neq 3$.
A comparison with similar methods in molecular dynamics is traced.
\end{abstract}
 
\section{Introduction}
 
Since the very gestation of the Newtonian mechanics \cite{Newton}, severe
criticism of its concepts of absolute space and time has proliferated. The
criticism concerning absolute space, as the tool determining the privileged
inertial frames is best expressed by Mach's principle \cite{Mach}: inertial
frames are an imprint of the matter content of the universe. According to
Leibniz \cite{Leibniz}, time is nothing more than the sequence of events.
Some of these ideas were incorporated in the general theory of relativity.
There, time is not absolute and Mach's principle is satisfied at least for
closed universes \cite{LBK,LBKB}. The generic idea of timelessness of both
a non-relativistic and a general relativistic complete universe was analyzed
recently by Barbour \cite{Barbour}.
 
In this generic framework there were also several attempts to incorporate
Mach's principle and Leibniz's concept of time in dynamical theories at
the non-relativistic level. Newtonian absolute space is eliminated from
such models and in some of them absolute time too, by implementing
time reparametrization invariance. However, in contrast with general
relativity, these models retain absulute simultaneity.
The study of such models dates back to the early attempts of Rei\ss ner
\cite{Rei1,Rei2}, who has considered the dynamics of the system of
$N$ interacting particles which depend on $N(N-1)/2$ mutual distances.
 
In the model developed by Lynden-Bell \cite{LB} both the kinetic and the
potential term in the Lagrangian are invariant under {\it time-dependent}
translations and rotations. This theory, as described by Lynden-Bell and
Katz \cite{LBK} is equivalent with Newtonian mechanics in a universe of zero
net angular momentum about its barycenter, provided the latter is in uniform
motion. There exists a strong evidence for the assumption of vanishing
angular momentum of the universe, given by the analysis of the cosmic
microwave background radiation \cite{BJS,H,CH}, in support to the validity of
the theory. Lynden-Bell's model describes an $N-$body system of interacting
particles in terms of purely relative quantities. No overall
(time-dependent) rotation of the frame about the barycenter or translational
motion of the barycenter has any effect on the action functional or the
equations of motion derived from it. This happens because no other distances
and velocities appear in the action than the distances between pairs of
particles and the differences of angular velocities of the lines joining
these pairs of particles, respectively. General relativistic extensions of
these ideas were discussed by Lynden-Bell, Katz and Bi\v {c}ak \cite{LBKB}.
In Lynden-Bell's model, however, time is absolute.
 
The{\it \ fully relational dynamical models} introduced by Barbour and
Bertotti \cite{BB1,BB2} are more general than Lynden-Bell's theory in
bearing invariance with respect to the whole Leibniz group \cite{BB1}
(including a monotonic redefinition of the time):
\begin{eqnarray}
x^{a^{\prime }} ={\cal R}_{\ b}^{a^{\prime }}(t)x^{b}+\Lambda ^{a^{\prime
}}(t)  \nonumber \\
t^{\prime } =t^{\prime }(t)\ .  \label{Leibniz}
\end{eqnarray}
Here ${\cal R}$ is an orthogonal matrix and $\Lambda ^{a^{\prime }}$ are
arbitrary translation parameters. As a consequence of employing an arbitrary
time parameter, the Lagrangian for a system of $N$ particles becomes a
homogeneous function of degree one in the ''velocities'' $(dq_{i}/dt^{\prime
},\ dt/dt^{\prime })$, implying via Euler's theorem a vanishing Hamiltonian
for the system \cite{Lanczos}. As an immediate consequence of this and the
invariance (\ref{Leibniz}) seven constants of the motion identically vanish.
These are the energy, the linear and angular momenta. Barbour and Bertotti
start from ''the relative motion of the universe treated as a single entity
and then recover the motion of subsystems within the background provided by
the universe at large''.The particular product Lagrangian proposed in \cite
{BB1} although being able to (qualitatively) reproduce some general
relativistic effects, like the perihelion shift and the Lense-Thirring
effect, leads also to unpleasant consequences: anisotropic masses (in
contradiction with experiment), time-dependent gravitational constant and a
violation of the Birkhoff theorem. In their second paper, Barbour and
Bertotti \cite{BB2} have proposed a more generic framework for deriving {\it
intrinsic dynamics}, relying on the concept of {\it intrinsic differential. }
Both Galilei- and Lorentz-invariance could be derived for particular
subsystems of the whole Leibniz-invariant universe.
 
As the theories of Barbour and Bertotti share with the theory of general
relativity the property of possessing both the Hamiltonian and momenta
constraints, the study of these models, particularly the issues concerning
quantization, can give an insight of how such theories should be handled in
a quantization process. Particularly interesting are the solutions given to
conceptual and interpretational difficulties by Barbour and Smolin \cite{BS1},
where an explicit solution is presented for the case of three particles in
one dimension. In a second paper \cite{BS2}, Barbour and Smolin introduce
the concept
of {\it variety }in order to provide a (non-local) description of dynamical
systems without any reference to elements lying outside the domain of
valability of these theories. The Barbour-Bertotti theories are among the
toy-models on which the applicability of their new concept and principle are
checked. A group-theoretical quantization of the Barbour-Bertotti theories
was done by Rovelli \cite{Rovelli}, eliminating any concept of time. In
contrast, an approach relying on a quantization in terms of an intrinsic
time was performed by Gambini and Mora \cite{GM}. Arguing that time
is a manifestation of dynamics, Barbour arrives to the notion of timelessness
of quantum gravity \cite{Barbour} and presents an approach to recover time
from a static wavefunction of the universe \cite{Barbour2}.
We will not enter in the details of the quantization process here. Our aim
in this and the subsequent companion paper \cite{GMcK} is to
analyze several aspects related to the geometric structure of these theories.
 
In what follows, by a Barbour-Bertotti type theory we mean a nonrelativistic
dynamical theory obeying the Hamiltonian, momenta and angular momenta
constraints. In Sec. 2 we start from the Newtonian description of the
interacting $N-$particle system, imposing the above constraints. Then,
by solving the constraints and eliminating the Lagrange-multipliers we obtain
the reduced Lagrangian, homogeneous in the velocities.
 
This action determines a Riemannian metric, in a form of a line element of
some Riemannian space. Extremizing the action is equivalent with finding the
geodesic motions. These issues are discussed in Sec. 3.
Although the particular form of the Lagrangian differs from that given
in \cite{BB1}, it has a product structure, in accordance with the general
framework settled in \cite{BB1,BB2}.
What we recover is the Jacobi principle for the constrained $N-$particle
system, as expected from the argumentation in \cite{Barbour}.
General relativity, in the form given by Baierlein, Sharp and Wheeler
\cite{BSW} resembles also a Jacobi principle rather than a parametrized
particle dynamics \cite{Barbour}.
 
In Sec. 4. we study the geometry of the reduced space for the generic case of
$N\geq 3$ particles. (For $N=1$ there is no relative distance at
all. For $N=2$, as the Lagrangian is homogeneous of degree one in the
single relative velocity, the Euler-Lagrange equation is trivially
satisfied and there is again no relative dynamics \cite{BB1}.) For
$N\geq 3$ we find that the metric is degenerate and the degeneracy
directions are related to the generators of translations and of rotations in
the physical space. This is again in accordance with \cite{BB1,BB2}, where
dynamics is defined on the space of orbits. We compute the curvature scalar
of the $3N-6$ dimensional reduced space (the space of orbits) by means of a
generalized Gauss equation, deduced here. The curvature scalar is expressed
in terms of the principal moments of inertia of the system. This
allows us to conclude that the line configurations are curvature
singularities for $N\neq 3$.
We will investigate the $N=3$ case in detail in the companion paper
\cite{GMcK}.
 
There are certain connections of our work with the analysis of the
$N-$body problem by geometric methods in molecular dynamics. A
recent summary of this topic was given by Littlejohn and Reinsch
\cite{LR}. We trace a comparison with their results in Sec. 5.
 
There are several type of indices appearing in our formulae.
The coordinates of the configuration space are labeled by a particle index,
denoted by a capital letter running from $1$ to $N$, and a Cartesian
coordinate index of the physical space (abstract index), denoted by a Latin
letter, which takes the values $1,2,3$. We adopt the convention to write the
summation over particles explicitly and apply the summation convention for
generic coordinate indices. The (Latin) indices of the Kronecker delta
$\delta_{ab}$ and of the totally antisymmetric Levi-Civita symbol
$\epsilon_{abc}$ are raised when necessary in order the summation convention
to apply. Kinematical and dynamical quantities written in a system
of coordinates originating in the center of mass carry a distinguishing zero
subscript, like $\mathop{x}\limits_{\circ }{}^{Ac}$. There is one exception
over this rule: the principal moments of inertia $I_{\mu }$ are written
without the zero subscript. Greek letters label coordinates in the system of
principal axes of inertia, with the origin in the center of mass. The
summation over these indices will be also explicitly indicated.
 
\section{The reduced Lagrangian of the Barbour-Bertotti model}
 
We start from the canonical form of the action characterizing a system of $N$
interacting particles:
\begin{equation}
S[x,p]=\int dt^{\prime }\left( \sum_{A=1}^{N}p_{Aa}\frac{dx^{Aa}}{dt^{\prime
}}-H\right) \ .  \label{actioncan}
\end{equation}
where $\ x^{Aa},\ dx^{Aa}/dt^{\prime }$ and $p_{Aa}$ are the coordinates,
velocity and momenta components, respectively of the particle with mass $
m_{A},$ and the Hamiltonian is given by:
\begin{equation}
H=\sum_{A=1}^{N}{\frac{1}{2m_{A}}}p_{Aa}p_{Ab}\delta ^{ab}+V\ .  \label{Ham}
\end{equation}
The potential term $V$ is a superposition of the two-particle potentials $
V_{AB},$ which depend only on the relative separation between particles $A$
and $B$:
\begin{equation}
V=\sum_{A<B}V_{AB}\left( \mid {\bf r}^{A}-{\bf r}^{B}\mid \right) \ ,
\label{Pot}
\end{equation}
${\bf r}^{A}=\{x^{Aa}\}$ being the position vector of particle $A.$
 
The Hamiltonian (\ref{Ham}) is time-independent, implying a conserved
energy. The total linear and angular momenta
\begin{equation}
P_{a}=\sum_{A=1}^{N}p_{Aa}\ ,\qquad L_{a}=\sum_{A=1}^{N}\epsilon _{ab}^{\ \
c}x^{Ab}p_{Ac}
\end{equation}
are also conserved:
\begin{equation}
\left\{ P_{a},H\right\} =0\ ,\qquad \left\{ L_{a},H\right\} =0\ .
\end{equation}
The above constants of the motion form a closed algebra:
\begin{equation}
\left\{ P_{a},P_{b}\right\} =0\ ,\quad \left\{ L_{a},L_{b}\right\} =\epsilon
_{ab}^{\ \ c}L_{c}\ ,\quad \left\{ P_{a},L_{b}\right\} =\epsilon _{ab}^{\ \
c}P_{c}\
\end{equation}
The last group of the above Poisson brackets bears the message that the
total momenta transform as vectors under rotations.
 
The Barbour-Bertotti theory is obtained from the above Newtonian theory by
imposing the constraints\footnote{
Without imposing $H=0$ and when $V_{AB}$ are the Newtonian gravitational
potentials, the Lynden-Bell theory \cite{LB} is recovered.}:
\begin{equation}
P_{a}=0\ ,\quad L_{a}=0\ ,\quad H=0\ .  \label{constraints}
\end{equation}
Among them, the condition $P_{a}=0$ is fulfilled by any system if the origin
of the frame is chosen in the center of mass. The other two constraints do
not hold in generic situations, for example in the case of an interacting
system of harmonic oscillators, where the energy is positive definite. For
attractive forces $H=0$ is the marginally bounded situation.
 
We add to the canonical action (\ref{actioncan}) the linear and angular
momenta constraints (\ref{constraints}) with Lagrange multipliers$\ {\cal N}
^{a}$ and ${\cal M}^{a}$. By replacing the time $t^{\prime }$ with a
labeling parameter $t$ \cite{Lanczos}, the Hamiltonian also gets a
multiplier ${\cal N}$:
\begin{equation}
S[x,p;{\cal N},{\cal N}^{a},{\cal M}^{a}]
=\int dt\left(\sum\limits_{A=1}^{N}p_{Aa}\dot{x}^{Aa}
\!-\!{\cal N}H
\!-\!{\cal N}^{a}P_{a}
\!-\!{\cal M}^{a}L_{a}\right)
\ .  \label{actionmult}
\end{equation}
(The notation $\dot{x}^{Aa}=dx^{Aa}/dt$ was introduced.) Variations of this
action as a functional of the multipliers generate the constraints.
Variations of (\ref{actionmult}) regarded as functional
of the coordinates and momenta give the equations of motion, equivalent with
the Newtonian equations for a system satisfying the constraints.
 
Next we proceed with the reduction of the action by solving the constraints
and expressing the Lagrange multipliers as functions of configuration data.
First by varying $p_{Aa}$ in (\ref{actionmult}) we find the relation between
the velocities and linear momenta:
\begin{equation}
\dot{x}^{Aa}={\cal N}{\frac{p_{Ab}}{m_{A}}}\delta^{ab}
+{\cal N}^{a}-{\cal M}_{\ b}^{a}x^{Ab}\ .  \label{vp}
\end{equation}
Here we have introduced the dualized form of ${\cal M}^{a}$, defined by the
relations:
\begin{equation}
{\cal M}_{ab}=\epsilon _{abc}{\cal M}^{c}\ ,\qquad {\cal M}^{a}={\frac{1}{2}}
\epsilon ^{abc}{\cal M}_{ab}\ .
\end{equation}
The unusual  relation (\ref{vp}) between the velocities and momenta
is consequence of the constrained nature of the
system (\ref{actionmult}). From Eq. (\ref{vp}) we express the momenta as:
\begin{equation}
p_{Aa}={\cal N}^{-1}m_{A}
\left( \dot{x}^{Ab}\delta_{ab}
-{\cal N}_{a}
+{\cal M}_{ab}x^{Ab}\right) \ .  \label{pv}
\end{equation}
The total linear momentum of the system is found by summing up the above
momenta over all particles:
\begin{equation}
P_{a}={\cal N}^{-1}M
\left( \dot{x}^{b}\delta_{ab}
-{\cal N}_{a}
+{\cal M}_{ab}x^{b}\right) \ .  \label{P}
\end{equation}
Here $M=\sum_{A=1}^{N}m_{A}$ is the total mass of the system, $
x^{a}=\sum_{A=1}^{N}m_{A}x^{Aa}/M$ and $\dot{x}_{a}$ are the coordinates and
the velocities of the center of mass, respectively.
 
In a similar manner by summing up over all particles the angular momenta
$l_{Aa}=\epsilon _{ab}^{\ \ c}x^{Ab}p_{Ac}$, the total angular momentum
emerges:
\begin{equation}
L_{a}={\cal N}^{-1}\left( l_{a}-I_{ab}{\cal M}^{b}-M{\cal N}
_{ab}x^{b}\right) \ .  \label{L}
\end{equation}
In the above expression we have introduced the velocity-based total angular
momentum:
\begin{equation}
l_{a}=\epsilon _{abc}\sum_{A=1}^{N}m_{A}x^{Ab}\dot{x}^{Ac}\ ,
\end{equation}
the tensor of inertia:
\begin{equation}
I_{gh}=\epsilon _{\ cg}^{l}\epsilon _{ldh}\sum_{A=1}^{N}m_{A}x^{Ac}x^{Ad}\
\end{equation}
and the dualized Lagrange-multiplier ${\cal N}_{ab}$ related to
${\cal N}^{a}$ as:
\begin{equation}
{\cal N}_{ab}=\epsilon _{abc}{\cal N}^{c}\ ,\qquad {\cal N}^{a}={\frac{1}{2}}
\epsilon ^{abc}{\cal N}_{ab}\ .
\end{equation}
 
By virtue of the constraints (\ref{constraints}) the equations (\ref{P}) and
(\ref{L}) form an inhomogeneous linear algebraic system in the multipliers
${\cal N}^{a}$ and ${\cal M}^{a}$. We transform this system to an equivalent
form by expressing the multipliers ${\cal N}_{a}$ from (\ref{P}) then
inserting them into (\ref{L}) obtaining:
\begin{equation}
{\cal N}_{a}={\cal M}_{ab}x^{b}+\dot{x}^{b}\delta_{ab}
\ ,\qquad
 \mathop{I}\limits_{\circ }{}_{ab}{\cal M}^{b}
=\mathop{l}\limits_{\circ }{}_{a}\ .
\label{system}
\end{equation}
Here
\begin{eqnarray}
\mathop{x}\limits_{\circ }{}^{Aa} = x^{Aa}-x^{a}=x^{Aa}-\frac{1}{M}
\sum_{A=1}^{N}m_{A}x^{Aa}  \label{x0} \\
\mathop{l}\limits_{\circ }{}_{a} =\epsilon _{abc}\sum_{A=1}^{N}m_{A}
\mathop{x}\limits_{\circ }{}^{Ab}\mathop{\dot x}\limits_{\circ }{}^{Ac} \\
\mathop{I}\limits_{\circ }{}_{gh} =\epsilon _{\ cg}^{l}\epsilon
_{ldh}\sum_{A=1}^{N}m_{A}\mathop{x}\limits_{\circ }{}^{Ac}\mathop{x}\limits
_{\circ }{}^{Ad}  \label{intens}
\end{eqnarray}
are the coordinates, the velocity-based angular momentum and the inertia
tensor in the center of mass frame, respectively.
 
There is a solution of the second equation (\ref{system}) only when the
tensor of inertia in the center of mass frame is nondegenerate
\footnote{
One's restricted ability in expressing the Lagrange multipliers in terms
of the configuration data arises also in the thin sandwich conjecture
\cite{BSW} of general relativity. The local solvability of that problem
for a large class of initial data was proved in \cite{BaFo}. Recently
results on the existence of solutions of the generalized
thin-sandwich problem were also obtained \cite{Giulini}.}
.  A careful study shows that a
degenerate inertia tensor would only occur in the case of all $N$ particles
being along a line (giving a rank $2$ inertia tensor) or in the unphysical
situation of all particles at the same point (rank $0$ inertia tensor). The
inverse inertia tensor $(\mathop{I}\limits_{\circ }{}^{-1})^{ab}$ is well
defined for all other configurations of the particles. Thus, modulo the
above exceptional cases, the multipliers ${\cal M}^{a}$ are:
\begin{equation}
{\cal M}^{a}=(\mathop{I}\limits_{\circ }{}^{-1})^{ab}\mathop{l}\limits
_{\circ }{}_{b}\ .  \label{Ma}
\end{equation}
If we insert the expressions of the multipliers ${\cal N}^{a}$ and ${\cal M}
^{a}$ from (\ref{system}) and (\ref{Ma}) in the expression of the momenta
(\ref{pv}) we obtain:
\begin{equation}
p_{Aa}={\cal N}^{-1}m_{A}
(\mathop{\dot x}\limits_{\circ }{}^{Ab}\delta_{ab}
+{\epsilon}_{abc}
(\mathop{I}\limits_{\circ }{}^{-1})^{cd}
 \mathop{l}\limits_{\circ }{}_d
 \mathop{x}\limits_{\circ }{}^{Ab})\ .
\label{momentafin}
\end{equation}
 
We have solved the linear and angular momenta constraints, thus the
terms containing the multipliers ${\cal N}^{a}$ and ${\cal M}^{a}$
can be dropped from the action (\ref{actionmult}). By inserting the
expression of the momenta (\ref{momentafin}), the remaining
Liouville form and lapse-Hamiltonian term become:
\begin{eqnarray}
\sum_{A=1}^{N}p_{Aa}\dot{x}^{Aa}
=2{\cal N}^{-1}\left( \mathop{T}\limits_{\circ }{}
-\frac{1}{2}\mathop{I}\limits_{\circ }{}_{ab}{\cal M}^{a}{\cal M}^{b}\right)
\\
{\cal N}H ={\cal N}^{-1}\left( \mathop{T}\limits_{\circ }{}-{\frac{1}{2}}
\mathop{I}\limits_{\circ }{}_{ab}{\cal M}^{a}{\cal M}^{b}\right) +{\cal N}V\
.
\end{eqnarray}
Here $\mathop{T}\limits_{\circ }{}$ is the kinetic energy in the center of
mass frame:
\begin{equation}
\mathop{T}\limits_{\circ }{}={\frac{1}{2}}\sum_{A=1}^{N}m_{A}
\mathop{\dot x}\limits_{\circ }{}^{Aa}
\mathop{\dot x}\limits_{\circ }{}^{Ab}
\delta_{ab}
\ .
\end{equation}
The action is now a functional on the configuration space and
function of the lapse alone:
\begin{equation}
S[x;{\cal N}]=\int dt{L}
\ ,\quad
{L}(x,\dot{x};{\cal N})={\cal N}^{-1}
\biggl(\mathop{T}\limits_{\circ }{}-{\frac{1}{2}}
\mathop{I}\limits_{\circ}{}_{ab}{\cal M}^{a}{\cal M}^{b}\biggr)-{\cal N}V
\ .  \label{actionn}
\end{equation}
The variation with respect to ${\cal N}$ gives an equation which we use
to express the lapse:
\begin{equation}
{\cal N}=\sqrt{\frac{\mathop{T}\limits_{\circ }{}-\frac{1}{2}\mathop{I}
\limits_{\circ }{}_{ab}{\cal M}^{a}{\cal M}^{b}}{-V}}=\sqrt{\frac{\mathop{T}
\limits_{\circ }{}-\frac{1}{2}({\mathop{I}\limits_{\circ }{}^{-1}})^{ab}
\mathop{l}\limits_{\circ }{}_{a}\mathop{l}\limits_{\circ }{}_{b}}{-V}}\ .
\end{equation}
By eliminating all multipliers, the task of reducing the action is
completed:
\begin{equation}
S[x]=\int\! dt{L}_{RED}
\ ,\quad
{L}_{RED}(x,\dot{x})=2\sqrt{-V\left(\mathop{T}\limits_{\circ }{}
\!-\!{\frac{1}{2}}({\mathop{I}\limits_{\circ }{}^{-1}})^{ab}
\mathop{l}\limits_{\circ }{}_{a}
\mathop{l}\limits_{\circ }{}_{b}\right) }\ .  \label{action}
\end{equation}
We have obtained the Jacobi Principle version of Lynden-Bell's purely
relative Lagrangian with the energy set to zero.
 
\section{The induced geometry}
 
The action (\ref{action}) obtained by completing the reduction
process (for all noncollinear configurations) is homogeneous
of degree one in the velocities. Therefore the
integrand can be regarded as a line element in some Riemannian space.
In order to make this manifest, we write the reduced Lagrangian
(\ref{action}) in two alternative forms:
\begin{eqnarray}
{L}_{RED}(x,\dot{x})
&&=\sqrt{-2V(x)\left[ \mathop{G}\limits_{\circ}{}_{AaBb}(x)
\mathop{\dot x}\limits_{\circ }{}^{Aa}
\mathop{\dot x}\limits_{\circ }{}^{Bb}\right] }
\nonumber\\
&&=\sqrt{-2V(x)\left[ G_{AaBb}(x)\dot{x}{}^{Aa}\dot{x}
{}^{Bb}\right] }\ ,  \label{Lred}
\end{eqnarray}
where $\mathop{G}\limits_{\circ }{}_{AaBb}$ and $G_{AaBb}$ are Riemannian
metrics defined as:
\begin{eqnarray}
\mathop{G}\limits_{\circ }{}_{AaBb} = m_{A}\delta _{AB}\delta
_{ab}-m_{A}m_{B}(\mathop{I}\limits_{\circ }{}^{-1})^{ef}\epsilon
_{eca}\epsilon _{fdb}\mathop{x}\limits_{\circ }{}^{Ac}\mathop{x}\limits
_{\circ }{}^{Bd}\   \label{G0} \\
G_{AaBb} =\mathop{G}\limits_{\circ }{}_{AaBb}
-\frac{m_{A}m_{B}}{M}\delta _{ab}\ .  \label{G}
\end{eqnarray}
Without changing the notations, from now on an index pair $Aa$ is
viewed as a single index running from $1$ to $3N$. Though the
reduction process at the Lagrangian level was accomplished, the
configuration space is still $3N$-dimensional. It will be one of
the tasks of the next section to show that the $3N$-dimensional
configuration space is superfluous and the $(3N-6)$-dimensional
space
of orbits of the Leibniz group is better suited as a configuration
space.
 
The metric has the simpler form (\ref{G0}) in the center of mass
frame, however this expression is covariant only under rotations in
the physical space. In contrast, the expression (\ref{G}), containing
an additional term, is covariant under both translations and rotations
in the physical space. It can be brought into a completely generic form
by inserting the expressions (\ref{x0}) of the coordinates
$\mathop{x}\limits_{\circ }{}^{Ac}$ in (\ref{G}) and in
$(\mathop{I}\limits_{\circ }{}^{-1})^{df}$ appearing there.
 
The geometry underlying the physical motions is characterized by the
conformally scaled (Jacobi) metric $-2V(x)G_{AaBb}$.
Extremizing $L_{RED}$ is equivalent with looking for the
geodesic motions associated with this metric. Dynamics arise from
geometry. To our knowledge it was
Synge \cite{Synge} who first applied this viewpoint in a discussion
of conservative systems in the framework of a
"geometro-dynamical theory of the manifold of configurations".
A recent discussion of the Jacobi metric for simple dynamical systems
\cite{Szydlowski} concentrates on the
singular regions at $E=V$ (at $V=0$ when specialized to our case).
 
As $V(x)$ varies from case to case, in the remaining part of the paper
we investigate the Riemannian metric (\ref{G}).
Formally, the metric $G_{AaBb}$ is characterizing the free
motions pertinent to the $V=const$ case.
 
At the end of this section we stress that the elimination of the
multiplier $\cal{N}$ from the action (\ref{actionn}) implies a choice
of time. This is because the elimination was carried out by employing
$\delta S[x;\cal{N}]/\delta \cal{N}$ which is the Hamiltonian
constraint $H=T+V=E=0$. However in a relational theory no time exists a
priori, in consequence the energy conservation equation is not giving the
speed of the system on its trajectory in the configuration space, as
usually. Rather it defines a unic time $t$ for all subsystems \cite{Barbour}.
 
\section{The reduction to the space of orbits}
 
For infinitesimal translations and rotations
$dx^{Aa}=d\xi^{a}+\epsilon _{\ bc}^{a}x^{Ab}d\eta ^{c}$
we have $G_{AaBb}dx^{Aa}=0\ .$
Therefore the metric is degenerated, the directions of degeneracies
being the vector flows in the configuration space induced by
translations and rotations in the physical space.
 
With respect to the flat metric
\begin{equation}
g_{AaBb}=m_{A}\delta _{AB}\delta _{ab}\ ,\qquad g^{AaBb}={\frac{1}{m_{A}}}
\delta ^{AB}\delta ^{ab}\ ,  \label{Gflat}
\end{equation}
the generators of translations $\xi _{(i)}^{Bb}=\delta _{i}^{b}$
and of rotations $\eta _{(j)}^{Bb}=\epsilon _{\ cj}^{b}x^{Bc}$
do not form an orthogonal set:
\begin{eqnarray}
\sum_{A,B=1}^{N}g_{AaBb}\xi_{(i)}^{Aa}\xi_{(j)}^{Bb}
=M\delta_{ij}
\nonumber \\
\sum_{A,B=1}^{N}g_{AaBb}\xi_{(i)}^{Aa}\eta_{(j)}^{Bb}
=\epsilon_{icj}x^{c}
\\
\sum_{A,B=1}^{N}g_{AaBb}\eta_{(i)}^{Aa}\eta_{(j)}^{Bb}
=I_{ij} \ ,\nonumber
\end{eqnarray}
excepting the case when the coordinate axes originate in the
center of mass
($x^c=0$ and $I_{ij}={\mathop{I}\limits_{\circ }}{}_{ij}$)
and they are chosen along the principal axes of the inertia tensor
(where ${\mathop{I}\limits_{\circ }}{}_{\mu\nu}=
I_{\mu }\delta _{\mu \nu }$,
$I_{\mu }$ denoting the principal moments of inertia).
In such a system an orthonormal set of six generators with
respect to the flat metric (\ref{Gflat}) is given by:
\begin{equation}
z_{(\mu )}^{B\beta}={\frac{1}{\sqrt{M}}}\delta _{\mu }^{\beta}
\ ,\qquad
w_{(\nu )}^{B\beta}={\frac{1}{\sqrt{I_{\nu }}}}
                      \epsilon _{\ \gamma\nu }^{\beta}
                      \mathop{x}\limits_{\circ }{}^{B\gamma}
\ .  \label{generators}
\end{equation}
The generators (\ref{generators}) correspond to translations
along, and rotations about the principal axes of inertia,
originating in the center of mass. The dual set of these
generators is:
\begin{equation}
u_{A\alpha}^{(\mu )}={\frac{m_{A}}{\sqrt{M}}}\delta _{\alpha}^{\mu }
\ ,\qquad
v_{A\alpha}^{(\nu )}={\frac{m_{A}}{\sqrt{I_{\nu }}}}
                     \epsilon _{\ \alpha \gamma}^{\nu }
                     \mathop{x}\limits_{\circ }{}^{A\gamma}
\ .  \label{dualgenerators}
\end{equation}
 
The expressions of the generators in generic frames are obtained
by the tensor transformation law:
\begin{eqnarray}
z_{(\mu )}^{Bb}=\sum_{C\gamma}z_{(\mu )}^{C\gamma}
                \frac{\partial{\mathop{x}\limits_{\circ}{}^{Bb}}}
                     {\partial{\mathop{x}\limits_{\circ}{}^{C\gamma}}}
\ ,\qquad
w_{(\nu )}^{Bb} = \sum_{C\gamma}w_{(\nu )}^{C\gamma}
                \frac{\partial{\mathop{x}\limits_{\circ}{}^{Bb}}}
                     {\partial{\mathop{x}\limits_{\circ}{}^{C\gamma}}}
\nonumber \\
u^{(\mu )}_{Aa}=\sum_{C\gamma}u^{(\mu )}_{C\gamma}
                \frac{\partial{\mathop{x}\limits_{\circ}{}^{C\gamma}}}
                     {\partial{\mathop{x}\limits_{\circ}{}^{Aa}}}
\ ,\qquad
v^{(\nu )}_{Aa} = \sum_{C\gamma}w^{(\nu )}_{C\gamma}
                \frac{\partial{\mathop{x}\limits_{\circ}{}^{C\gamma}}}
                     {\partial{\mathop{x}\limits_{\circ}{}^{Aa}}}
\ .  \label{generators_gen}
\end{eqnarray}
 
We conclude that $G_{AaBb}$ is the metric in the $(3N-6)$ dimensional space
of orbits of the group of translations and rotations, rigged by the generators
(\ref{generators_gen})
\footnote{ For a generic discussion of a rigged $(n-m)$ dimensional manifold
in an $n-$dimensional manifold endowed with a linear connection see Schouten
\cite{Schouten}}
:
\begin{equation}
G_{AaBb} =g_{AaBb}-\sum_{\mu }u_{Aa}^{(\mu )}u_{Bb}^{(\mu )}-\sum_{\nu
}v_{Aa}^{(\nu )}v_{Bb}^{(\nu )}
\ .
\end{equation}
 
We define
\begin{equation}
P_{Bb}^{Aa} =\delta _{B}^{A}\delta _{b}^{a}-\sum_{\mu }z_{\ (\mu
)}^{Aa}u_{Bb}^{(\mu )}-\sum_{\nu }w_{(\nu )}^{Aa}v_{Bb}^{(\nu )}\ .
\end{equation}
It is an easy exercise to verify that $P_{Bb}^{Aa}$ is a projection
operator
\begin{equation}
P_{Cc}^{Aa}P_{Bb}^{Cc}=P_{Bb}^{Aa}
\label{proj}
\end{equation}
to the subspace rigged by the generators:
\begin{equation}
P_{Bb}^{Aa}z_{\ (\mu )}^{Bb}=P_{Bb}^{Aa}u_{Aa}^{(\mu )}=0\ ,\qquad
P_{Bb}^{Aa}w_{\ (\nu )}^{Bb}=P_{Bb}^{Aa}v_{Aa}^{(\nu )}=0\ .
\label{rigging}
\end{equation}
 
The construction we have found is quite generic. Whenever one has a Riemannnian
manifold and a gauge group of isometries acting on it, the metric on the
quotient space is naturally given by the restriction of the initial metric to
the orthogonal complement to the gauge group.
We stress that none of $w_{(\nu )}^{Bb}$ and $v^{(\nu )}_{Aa}$ and neither
$G_{AaBb}$ are well-defined when any of the principal moments of inertia
vanish. Thus we have to limit the validity of the above construction to the
noncollinear configurations.
The reduced space is the quotient space $R^{3N}/E(3)$ where
$E(3)=R^3 \rtimes SO(3)$. By extending this manifold
to include the collinear
configurations either, we find a manifold with dimensionality $3N-6$ with
boundary for $N=3$ and without boundary for $N\ge 4$ \cite{LR}.
The case $N=3$ will be discussed in detail in the companion paper
\cite{GMcK}. For $N=4$ the manifold in discussion is homeomorphic with
$R^6$ \cite{LR2}.
 
We denote by $\nabla $ and $\tilde{\nabla}$ the connections
compatible with the metric $G$ and the flat metric $g$.
The respective covariant derivatives of an arbitrary vector field $q^{Aa}$
in the coordinate basis $\{ \frac{\partial}{\partial x^{Bb}} \}$
will be denoted by $\nabla_{Cc}q^{Aa}$ and $\tilde\nabla_{Cc}q^{Aa}$.
For each of the generators (\ref{generators}) we define one of the tensors:
\begin{eqnarray}
\chi _{AaBb}^{(\mu )}
=\sum_{E,F=1}^{N}P_{Aa}^{Ee}P_{Bb}^{Ff}\tilde{\nabla}_{Ee}u_{Ff}^{(\mu )}
\nonumber\\
\omega _{AaBb}^{(\nu)}=
\sum_{E,F=1}^{N}P_{Aa}^{Ee}P_{Bb}^{Ff}\tilde{\nabla}_{Ee}v_{Ff}^{(\nu )}
\ .  \label{chiom}
\end{eqnarray}
Pairs of indices $Aa$ of these tensors are raised and lowered with the flat
metric (\ref{Gflat}).
 
From straightforward algebra on Eq. (\ref{intens}) written
in the principal axis frame the following useful expression emerges:
\begin{equation}
\sum_{A=1}^{N}m_{A}\mathop{x}\limits_{\circ }{}^{A\mu }\mathop{x}\limits
_{\circ }{}^{A\nu }={\frac{\delta ^{\mu \nu }}{2}}\sum_{\rho }(1-2\delta
_{\mu \rho })I_{\rho }\ .  \label{sum}
\end{equation}
By use of the above relation, together with Eqs. (\ref{generators}),
(\ref{dualgenerators}) and (\ref{x0}), we compute the expressions for the
$\chi^{(\mu )}$ and $\omega^{(\nu )}$ tensors in the principal axis frame
in terms of the principal moments of inertia:
\begin{eqnarray}
\chi_{A\alpha B\beta}^{(\mu )}= 0
      \nonumber \\
\omega_{A\alpha B\beta}^{(\nu )}=
-\frac{m_B}{\sqrt{I_{\nu}}}
\epsilon_{\ \alpha\beta}^{\nu}
\Bigl(\delta_{AB}-{\frac{m_A}{M}}\Bigr)
      \nonumber \\
+
{\frac{m_Am_B}{\sqrt{I_{\nu}}}}
\sum_{\gamma\delta\rho}
\mathop{x}\limits_{\circ }{}^{B\gamma}
\mathop{x}\limits_{\circ }{}^{A\delta}
\Biggl[
\epsilon _{\ \alpha\rho}^{\nu }
\Bigl(\sum_{\sigma }
 {\frac{\epsilon _{\ \delta\sigma}^{\rho}
        \epsilon _{\ \beta \gamma}^{\sigma}}
       {I_{\sigma }}}\Bigr)
+
\epsilon _{\ \ \beta}^{\nu \rho}
\Bigl(\sum_{\sigma }
 {\frac{\epsilon_{\alpha\delta\sigma}
        \epsilon_{\ \rho\gamma}^{\sigma}}
       {I_{\sigma }}}\Bigr)
      \nonumber \\
-
\sum_{\lambda\mu\chi}
\epsilon _{\ \ \lambda}^{\nu\rho}
\Bigl(\sum_{\sigma }
 {\frac{\epsilon_{\alpha\delta\sigma}
        \epsilon_{\ \rho\mu}^{\sigma}}
       {I_{\sigma }}}\Bigr)
\Bigl(\sum_{\tau }
 {\frac{\epsilon_{\ \chi\tau}^{\lambda}
        \epsilon_{\ \beta\gamma}^{\tau }}
       {I_{\tau }}}\Bigr)
\Bigl(\sum_{G=1}^{N}m_{G}
\mathop{x}\limits_{\circ }{}^{G\mu}
\mathop{x}\limits_{\circ }{}^{G\chi}\Bigr)
\Biggr]
\ .  \label{chiomexpl}
\end{eqnarray}
The tensor transformation law gives the expression of
$\omega^{(\nu)}$ in a generic frame:
\begin{equation}
\omega_{AaBb}^{(\nu )}=
\sum_{CD\alpha\beta}
\omega_{C\alpha D\beta}^{(\nu )}
{\frac{\partial{\mathop{x}\limits_{\circ}{}^{C\alpha}}}
      {\partial{\mathop{x}\limits_{\circ}{}^{Aa}}}}
{\frac{\partial{\mathop{x}\limits_{\circ}{}^{D\beta}}}
      {\partial{\mathop{x}\limits_{\circ}{}^{Bb}}}}\ .
\end{equation}
Similarly, $\chi _{AaBb}^{(\mu )}=0$ holds in any frame.
 
The tensor $\chi _{AaBb}^{(\mu )}$ being symmetric (it vanishes) is the
extrinsic curvature of the hypersurfaces with normal vectors $u_{\mu }^{Aa}$.
By contrast, the tensor $\omega _{AaBb}^{(\nu )}$ is not hypersurface
orthogonal, being antisymmetric. It is the vorticity tensor for the vectors
$w_{\nu }^{Aa}$. This antisymmetry originates in the transformation properties
of $w_{\ (\nu )}^{Bb}$ under reflections: it behaves as an
axial (pseudo) vector rather then a polar (true) vector. The antisymmetry of
the tensor $\omega _{AaBb}^{(\nu )}$ complicates the relation between the
Riemann curvature tensors $\tilde{R}_{\ \ BbCcDd}^{Aa}$ (here vanishing) and
$R_{\ \ BbCcDd}^{Aa}$, associated to the metrics $g_{AaBb}$ and $G_{AaBb}$,
respectively.
 
We derive the required generalized Gauss equation as follows. Let $q^{Bb},\
r^{Cc}$ and$\ s^{Dd}$ be three arbitrary vectors and $p_{Aa}$ an arbitrary
one-form in the tangent and cotangent spaces of the space of orbits,
respectively. The two Riemann tensors are defined as:
\begin{eqnarray}
\sum_{B=1}^{N}\tilde{R}_{\ \ BbCcDd}^{Aa}\ q^{Bb}
=2\tilde{\nabla}_{[Cc}\tilde{\nabla}_{Dd]}q^{Aa}
\nonumber\\
\sum_{B=1}^{N}R_{\ \ BbCcDd}^{Aa}\ q^{Bb}
=2\nabla _{[Cc}\nabla _{Dd]}q^{Aa}\ .  \label{Riemanndef}
\end{eqnarray}
Evaluating the expression $\tilde{R}_{\ \ BbCcDd}^{Aa}\
p_{Aa}q^{Bb}r^{Cc}s^{Dd}$, we get the desired relation between the Riemann
curvature tensors of the metric $g$ on the $3N$ dimensional space and of the
metric $G$ on the rigged $(3N-6)$ dimensional space:
\begin{eqnarray}
\sum_{B,E,F,G,H=1}^{N}\!\tilde{R}_{\ \ FfGgHh}^{Ee}
P_{Ee}^{Aa}P_{Bb}^{Ff}P_{Cc}^{Gg}P_{Dd}^{Hh}\ q^{Bb} =
2\sum_{\nu }\omega_{[CcDd]}^{(\nu )}\left( {\ell }_{w_{(\nu )}}q^{Aa}\right)
\nonumber \\
+\sum_{B=1}^{N}\left[ R_{\ \ BbCcDd}^{Aa}
-2\sum_{\nu }\left( \omega_{[Cc}^{(\nu )\ Aa}\omega _{Dd]Bb}^{(\nu )}
-\omega _{[CcDd]}^{(\nu )}\omega_{Bb}^{(\nu )\ Aa}\right) \right] \ q^{Bb}
\ ,
\label{Riemann}
\end{eqnarray}
where ${\ell }_{w_{(\nu )}}q^{Aa}$ denotes the Lie-derivative ${\cal L}
_{w_{(\nu )}}$ of the vector $q^{Aa}$, projected to the reduced space:
\begin{equation}
{\ell }_{w_{(\nu )}}q^{Aa}=P_{Bb}^{Aa}{\cal L}_{w_{(\nu )}}q^{Bb}\ .
\end{equation}
The major inconvenience induced by the antisymmetric $\omega _{AaBb}^{(\nu
)} $ tensors in (\ref{Riemann}) is the presence of the derivative term of $
q^{Bb}$. For generic riggings such a term should be absorbed in the definition
of the Riemann tensor of the reduced space as described by Schouten \cite
{Schouten}. However in our case the situation is simpler. As
translations and rotations are symmetries of the system, the quantities
defined on the space of orbits (in particular the components of the tangent
vector $q^{Aa}$) should not depend on the parameters of
translations and rotations. In an adapted coordinate system the
Lie-derivatives are partial derivatives with respect to these parameters,
thus the Lie-derivative terms in (\ref{Riemann}) vanish. Because the
vanishing of a tensorial quantity is a invariant statement, this result
holds in arbitrary coordinate system as well. Therefore and by employing the
flatness of the metric $g$, the expression of the Riemannian curvature
tensor of the rigged space emerges purely in terms of the vorticity tensors $
\omega ^{(\nu )}$:
\begin{equation}
R_{\ \ BbCcDd}^{Aa}=2\sum_{\nu }\left( \omega _{[Cc}^{(\nu )\ Aa}\omega
_{Dd]Bb}^{(\nu )}-\omega _{CcDd}^{(\nu )}\omega _{Bb}^{(\nu )\ Aa}\right)
\ .  \label{Riemann1}
\end{equation}
This is the desired generalized Gauss equation.
 
The curvature scalar is then readily found:
\begin{equation}
R=\sum_{A,B,D=1}^{N}G^{BbDd}R_{\ \ BbAaDd}^{Aa}=-3\sum_{\nu }\omega
_{Bb}^{(\nu )\ Aa}\omega _{Aa}^{(\nu )\ Bb}\ .  \label{R}
\end{equation}
Here $G^{BbDd}$ denotes the inverse metric of the rigged space. A lengthy,
but straightforward computation, performed in the system of principal axes
of inertia, employing the explicit expressions of the vorticity tensors
(\ref{chiomexpl}) and the auxiliary relation
(\ref{sum}) has given the scalar curvature in terms of the
principal moments of inertia and the number of particles:
\begin{equation}
R=6(N-2)\sum_{\nu }{\frac{1}{I_{\nu }}}-{\frac{3}{2I_{1}I_{2}I_{3}}}
\Bigl( {\sum_{\nu }I_{\nu }}\Bigr) ^{2}\ .  \label{R1}
\end{equation}
$R$ being scalar, the above expression is independent of the particular
frame employed for its evaluation.
 
The concise form (\ref{R1}) allows one to
find out whether the manifold of orbits can be extended by the inclusion of
the alligned configurations. The metric $G_{AaBb}$ extended there by a
limiting process, is singular. The question arises, whether
these line configurations represent coordinate or real
singularities. Suppose that the system is passing through a sequence of
configurations ${\cal C}_{n}$ toward the line configuration ${\cal C}_{line}$
with all particles aligned on the $x$-axis. There $I_{1}=0$ while $
I_{2}=I_{3}=I$ are finite. The curvature scalar (\ref{R1}) behaves as
\begin{equation}
\lim_{{\cal C}_{n}\to {\cal C}_{line}}R=\frac{6(2N-5)}{I}+\frac{6(N-3)}{I_{1}
}\ .  \label{Rlim}
\end{equation}
Excepting the case $N=3$ the curvature scalar diverges together with the
vanishing of $I_{1}$. Thus the line configurations are curvature
singularities for $N\neq 3$.
 
\section{Connections with molecular dynamics}
 
It is instructive to compare our description with a geometric
treatment of the $N-$body problem in the language of fiber bundles, employed
in both classical and quantum molecular dynamics. A recent review of this
topic was given by Littlejohn and Reinsch \cite{LR}. In that approach
the translational degrees of freedom are subtracted from the very
beginning by the introduction of the $3N-3$ Jacobi coordinates.
When the configurations are not collinear, the remaining structure
is identified with a principal fiber bundle with the structure group
$SO(3)$. The orbits of this group are the fibers and the base space is
what we have called the reduced configuration space (shape
space in the language of \cite{LR}). Two kind of frames play an essential
role in this approach: the space frame, which is an externally prescribed
inertial frame, and the body frame, which can be attached to a flexible
body in many ways. This arbitrariness in the choice of the body frame is
a gauge freedom, any choice of gauge corresponding to a section of the
fiber bundle. A gauge potential $A_a^i$ is introduced. The decomposition
of the motion in rotational and vibrational part is gauge dependent,
however the curvature form,
\begin{equation}
B_{ab}^i=\frac{\partial A_b^i}{\partial q^a}
        -\frac{\partial A_a^i}{\partial q^b}
        -\epsilon^i_{\ jk}A_a^jA_b^k\ ,
\end{equation}
called also Coriolis tensor is gauge covariant. Here $\{q^a\}$ are
generic coordinates in shape space and the indices $a,b$ symbolize the
tensorial character of the Coriolis tensor with respect to changes in
the shape space coordinates, while indices $i,j,k$ denote the components
in some body frame. Thus the Coriolis tensor is a geometric object with
$3\times(3N-6)\times(3N-6)$ components.
 
A metric $G_{ab}$ in shape space is also introduced by means of the horizontal
contribution to the kinetic energy. By a Kaluza-Klein type decomposition the
Riemann tensor and curvature scalar associated to this metric are found,
respectively:
\begin{eqnarray}
R_{abcd}=\frac{1}{2}
     \left(B^i_{ab }\mathop{I}\limits_{\circ }{}_{     ij    }B^j_{cd}
          -B^i_{a[c}\mathop{I}\limits_{\circ }{}_{\mid ij\mid}B^j_{d]b}\right)
\label{RimLR}\\
R=\frac{3}{4}B^{iab}\mathop{I}\limits_{\circ }{}_{ij}B^j_{ab}\ .
\label{curvLR}
\end{eqnarray}
In the second formula the indices $a,b$ were raised by the inverse of the
(non-flat) metric $G_{ab}$.
 
Specializing to vanishing angular momentum leads to horizontal motions in the
principal fiber bundle. It is thus not surprising, that we have exploited a
somewhat simpler geometric structure in our analysis.
 
Our approach differs in many aspects, as we were discussing a purely
relational theory. We cannot start from the concept of space frame.
It would imply to monitor the evolution of the system with respect to some
preexisting inertial system, which is not the case.
By contrast, we have chosen a definite body frame for our computations in
Sec. 4. This is the principal axis frame.
 
We have imposed constraints on the dynamics by means of
Lagrange-multipliers, at the end of the reduction process arriving to
a Jacobi-type action (\ref{action}). The Lagrangian in \cite{LR},
without the kinetic energy of the center of mass and  specialized to
vanishing angular momentum (when the vertical contribution
to the kinetic energy about the center of mass vanishes) is
\begin{equation}
L(q,\dot q)=\frac{1}{2}G_{ab}\dot q^a\dot q^b-V(q) \ .
\label{LagLR}
\end{equation}
By writing the Jacobi principle version of this action with the energy set
to zero we can identify the metric $G_{ab}$ with our metric $G_{AaBb}$,
expressed in a coordinate system adapted to the six generators.
 
By comparing our formulae (\ref{Riemann1}) and (\ref{R})
for the Riemann tensor and curvature scalar with Eqs. (\ref{RimLR}) and
(\ref{curvLR}) respectively it is tempting to identify our vorticity tensor
$\omega^{(\nu)}_{AaBb}$ with the Coriolis tensor. For this purpose
first we write Eq. (\ref{curvLR}) in the principal axis frame:
\begin{equation}
R=3\sum_{\nu}\frac{\sqrt{I_{\nu}}B^{\nu ab}}{2}
             \frac{\sqrt{I_{\nu}}B^{\nu}_{ab}}{2}
\ .
\label{curvLRpaf}
\end{equation}
Second, we remark that unlike the Coriolis tensor, the vorticity tensors
are defined as three geometric object with $3N\times3N$ components.
However by their definition (\ref{chiom}) as projections we see that the
vorticity tensors live in the $(3N-6)-$dimensional reduced space
(shape space). When coordinates adapted to this space are chosen, they will
have equally $(3N-6)\times(3N-6)$ nonvanishing components.
Third, indices $Aa,Bb$ are raised with the (inverse) flat metric
$g^{AaBb}$, in contrast with the contravariant form of the Coriolis tensor
in (\ref{curvLRpaf}), where indices are raised with the nonflat metric $G^{ab}$.
But there is no difference in raising the indices of $\omega^{(\nu)}_{AaBb}$
with the nonflat metric $G^{AaBb}$ either. This is because the two metrics
differ only in terms homogeneous in the rigging vectors. When contracted with
the vorticity tensor (a projected object) these difference terms vanish
by virtue of Eqs. (\ref{rigging}).
 
Thus we identify the set of the three vorticity tensors $\omega^{(\nu)}$ with
$\sqrt{I_{\nu}}B^{\nu}/2$ (no summation). The advantage in employing the
vorticity tensors, which are defined on the (unreduced) configuration space
is that their indices can be raised and lowered with the flat metric $g$.

\section{Concluding Remarks}
 
The study of the Barbour-Bertotti theory describing the Newtonian $N-$
particle system constrained by the vanishing of all constants of the
motion revealed that a complete reduction at the Lagrangian level is
possible whenever the particles are not along a line.
 
The reduced space
emerged as the space of orbits of the Leibniz-group. Free motions are
geodesics with respect to the Riemannian metric $G_{AaBb}$
while motions characterized by $V\neq const$ are geodesics
of the conformally scaled metric $-2V(x)G_{AaBb}$. The vorticity
tensors of the generators of rotations were introduced and their
expression in terms of the pricipal moments of inertia was derived.
 
The Riemann tensor and the curvature scalar associated with the metric
$G_{AaBb}$ were given in terms of the vorticity tensors. The additional
expression of the scalar curvature in terms of the principal moments of
inertia allowed for a study of the collinear configurations. Curvature
singularities arise in these configurations, unless the number of
particles is three.
 
The Riemannian curvature tensor and the curvature scalar associated to
the conformally scaled metric $\hat G_{AaBb}=-2V(x)G_{AaBb}$, which
applies to the case $V\neq const$ can be found by applying the
relations among these objects for conformally related metrics
\cite{Wald}. For illustration we give here the curvature scalar $\hat R$
associated to the metric $\hat G$:
\begin{eqnarray}
\hat R=\frac{1}{-2V}
\Biggl\{ R&-&(3N-7)G^{AaBb}
\Biggl [\nabla_{Aa}\nabla_{Bb}\log (-V)
\nonumber \\
       &+&\left (\frac{3N}{4}-2\right )
          \nabla_{Aa}\log (-V)\nabla_{Bb}\log (-V)
\Biggr ]
\Biggr\}
\ .
\label{hatR}
\end{eqnarray}
For any definite choice of the potential $V$, such relations
can be further exploited.

\ack
 
The description of the Barbour-Bertotti model in Sec. 2 is based
on unpublished notes by Karel Kucha\v {r}. It is a pleasure to thank
him for suggesting the problem and for useful references.
The author is grateful for the warm hospitality of the relativity
group at the Physics Department of University of Utah
and for clarifying discussions with L\'aszl\'o Szabados and
Mitchell McKain in the early stages of this work.
The comments of the referees led to improvements in the
presentation and they are acknowledged.
This work was realized with the continued support of the NSF grant
PHY-9734871, OTKA grants W015087 and D23744, the E\"otv\"os Fellowship
and the Soros Foundation.

\end{document}